\documentclass{nature}
\usepackage{graphicx}

\title{Self-similarity of Dynamo Action in the Largest Cosmic Structures}

\author{Francesco Miniati$^{1}$ \& Andrey Beresnyak$^2$}

\begin{document}

\maketitle

\begin{affiliations}
 \item Physics Dept., ETH Zurich, Wolfgang-Pauli-Strasse 27, CH-8093 Switzerland
 \item Nordita, KTH Royal Institute of Technology and Stockholm University, SE-10691 Stockholm, Sweden
\end{affiliations}

\begin{abstract}
Massive galaxy clusters (GC) are filled with a hot, turbulent and magnetised intra-cluster medium (ICM). Still 
forming under the action of gravitational instability they grow in mass by accretion of supersonic flows. These 
flows partially dissipate into heat through a complex network of large scale shocks\cite{2000ApJ...542..608M},
while residual transonic flows 
create giant turbulent eddies and cascade\cite{2011A&A...529A..17V,2014ApJ...782...21M}. 
Turbulence heats the ICM\cite{2014Natur.515...85Z} and also amplifies magnetic energy by 
way of dynamo action\cite{schluter1950,2009ApJ...693.1449C,2008Sci...320..909R,B12a}.
However, fundamental properties of 
the pattern whereby gravitational energy turns kinetic, thermal, turbulent and magnetic remain unknown.
Here we find that the energy components of the ICM are ordered according to a permanent hierarchy,
in which the ratios of thermal to turbulent to magnetic energy densities remain virtually unaltered 
throughout the ICM history despite evolution of each individual  
component and the drive towards equipartition of turbulent dynamo.
Our results are based on a state-of-the-art, fully cosmological computational model of 
ICM turbulence\cite{2014ApJ...782...21M,2015ApJ...800...60M}, revealing that
an approximately constant efficiency of turbulence generation 
from gravitational energy that is freed during mass accretion.
The permanent character of this hierarchy reflects a new aspect of self-similarity in
cosmology\cite{1970PhRvD...1.2726H,1986MNRAS.222..323K,1997ApJ...490..493N,1999ApJ...524L..19M}, 
while its structure, consistent with current 
data~\cite{2001ApJ...547L.111C,2008A&A...483..699G,Bonafede2010,Govoni2010,2011A&A...529A..13K},
encodes information about the efficiency of turbulent heating and dynamo action.
\end{abstract}

The computational model captures the turbulent motions through a multi scale technique 
which employs six nested grids covering progressively larger volumes with correspondingly coarser resolution 
elements~\cite{2014ApJ...782...21M,2007JCoPh.227..400M}.
The finest grid resolves the virial volume of the GC with more than a billion uniform-size resolution 
elements and provide the necessary dynamic range to resolve the ICM turbulent cascade. The largest grid 
covers the chosen cosmological volume of 340 comoving Mpc on a side (comoving=partaking in the 
expansion of the universe; 1 Mpc $\approx$ 3 million ly). 
The intermediate grids allow to simultaneously follow with 
adequate accuracy the matter distribution outside the GC volume. 
The calculation starts with three grids and adds 
progressively finer grids as the Lagrangian volume of the GC shrinks 
under self-gravity. All six grids are in place 
at a time corresponding to 8 billion yr after Big Bang. At current time
(13.8 billion years after Big Bang) the simulated GC has a
 total virial mass of $1.3\times10^{15} M_\odot$.

Figure~\ref{GC:fig} shows a snapshot of the simulation
illustrating the cosmological context and the 
highly turbulent conditions of the flow inside the GC volume. 
The exquisite resolution across the GC volume allows us to accurately measure the time dependent 
statistical properties of structure formation driven ICM turbulence
including, in particular, the dissipation rate, $\epsilon_{turb}$,
the outer scale, $L$ and the velocity dispersion on that scale, $\langle(\delta u_L)^2\rangle^\frac{1}{2}$
(Methods). 
In the following we restrict our analysis to a region within 1/3 of the GC's virial radius, $R_{vir}$,
where $R_{vir}$ defines a region characterised by a mass over-density $\Delta_c\approx 100$
that has nominally reached dynamical equilibrium.
Our choice is motivated by the fact that at current epoch $R_{vir}/3\approx 1$~Mpc, i.e. defines a region 
most relevant for comparison with observations.

Hydrodynamic turbulence is dominated by the solenoidal component 
accounting for 60-90\% of the total kinetic energy\cite{2015ApJ...800...60M,BM15}.
Detailed analysis shows that  this component remains statistically 
homogenous and isotropic thus resembling Kolmogorov's cascade,
despite the presence of considerable structure in the ICM\cite{2015ApJ...800...60M}.
The dissipation of incompressible turbulence contributes to ICM heating along with 
shocks and adiabatic compression and to the growth of magnetic energy
by way of small scale dynamo action\cite{schluter1950,B12a} (Methods: 
Fig.~\ref{ICM_turb_diagram:fig} for cascade details).
The turbulent dissipation rate associated to the solenoidal component is estimated
from the numerical simulation data.
Because ICM turbulence is driven by various complex 
hydrodynamic mechanisms ultimately powered by the unsteady mass accretion process\cite{2015ApJ...800...60M}, the dissipation rate is highly changeable with time and 
exhibits non-monotonic variations by more than one order of magnitude\cite{BM15} (Figure~\ref{epsilon}c).
However, alongside with much complexity turbulent dissipations appears to also exhibit 
simplicity of behaviour.
This is shown in Figure~\ref{epsilon}a illustrating the time evolution of 
the fraction of thermal energy originating from turbulent dissipation.
In contrast to $\epsilon_{turb}$, this quantity remains remarkably constant during 
the GC lifetime, $\eta_{turb}\approx 0.3-0.4$, indicating that the efficiency 
of turbulence generation out of gravitational energy freed by mass accretion is approximately constant.
In addition, Figure~\ref{epsilon}b shows that the turbulence velocity dispersion at the outer scale 
normalised to the ICM sound speed, i.e. the turbulence Mach number
${\cal M}_{turb}=\langle(\delta u_L)^2\rangle^\frac{1}{2}/c_s$, also remains rather constant with time.
This shows that in the ICM the evolution of the turbulent kinetic energy and the 
thermal energy are closely related, consistent with the previous plot.
The value of ${\cal M}_{turb}$ can be understood as follows. 
If the generation of the bulk of the thermal energy, $E_{th}$,
is dominated by the last $\alpha=2-3$ eddy turnover times, $\tau_L=L/u_L$, then
$E_{th}\simeq \eta_{turb}^{-1}\int \rho \epsilon_{turb}\,dt \simeq (\alpha/3^{\frac{3}{2}}\eta_{turb}) \rho 
\langle(\delta u_L)^2\rangle$, 
where we have used the known relation $\epsilon_{turb}=(2/3C)^\frac{3}{2}
\langle(\delta u_L)^2\rangle^\frac{3}{2}/L$ with $C\approx 2$\cite{Landau:111625}.
It is straightforward to then see that the Mach number 
${\cal M}_{turb}\approx (\sqrt{3}/\alpha)^\frac{1}{2}(\eta_{turb}/0.37)^\frac{1}{2}$ 
which, for $\alpha=1.5-3$ 
ranges between 0.8 and 1.2 confirming the result in Figure~\ref{epsilon}.
It also follows that the ratio of thermal to turbulent kinetic energy is
\begin{equation}
\frac{E_{th}}{\frac{1}{2}\rho\langle(\delta u_L)^2\rangle} \approx \frac{2\alpha}{3^\frac{3}{2}}\eta_{turb}^{-1} 
\approx  \eta_{turb}^{-1}.
\end{equation}

Generation of magnetic field by small scale dynamo in a turbulent flow follows from standard theory.
In a high Reynolds number ($Re>10^3$) flow such as the ICM an initial seed of vanishing strength\cite{1994MNRAS.271L..15S,1997ApJ...480..481K,2011ApJ...729...73M,2012Natur.481..480G,2006MNRAS.370..319B,2009MNRAS.392.1008D} is amplified exponentially
at the rate $\gamma=\sqrt{\it Re}/30\tau_L$, where $\tau_L =L/\delta u_L$ is the eddy turnover time.
After a short while ($\propto Re^{-1/2}$) magnetic 
field stops growing below a characteristic Alfv\'en scale, 
$L_A\equiv v_A^3/C^\frac{3}{2}\epsilon_{turb}$,
where $v_A=B/\sqrt{4\pi\rho}$ is the Alfv\'en speed, due to the feedback action of magnetic tension\cite{schluter1950,B12a}. 
Magnetic energy continues to grow at the expenses of turbulent 
kinetic energy as $L_A$, marking the equipartition scale between 
kinetic and magnetic energy, shifts towards larger values\cite{schluter1950}. Growth, however, is now
proportional to the turbulent dissipation rate
instead of exponential with time. It is in this latter stage that the dynamo spends most of the time\cite{B12a}. 
Recent state-of-the-art numerical work finds that  for statistically isotropic and 
homogeneous turbulence, as found in the ICM\cite{2015ApJ...800...60M, BM15}, the efficiency of conversion of 
turbulent (kinetic) to magnetic energy is a universal number ca $C_E$= 4-5\%\cite{B12a}.

Therefore, the evolution of magnetic energy in the ICM can be expressed in terms of the
turbulence dissipation history as $E_B(t)= B^2/8\pi=C_E \int^t d\tau \rho\epsilon (\tau)$.
Combined with the above finding about $\eta_{turb}$, 
this leads to simple but significant expressions relating the
fundamental properties of magnetic field and and turbulence in the ICM.
In fact, since turbulence dissipation contributes a constant fraction, 
$\eta_{turb}\approx 1/3$, of ICM thermal 
energy,  the ratio $\beta_{plasma}$ of thermal pressure to magnetic energy can be written as
\begin{equation} 
\beta_{plasma}\equiv\frac{P_{gas}}{B^2/8\pi}=\frac{\eta_{turb}^{-1}(\gamma-1)}{C_E}
=40\left(\frac{\eta_{turb}}
{1/3}\right)^{-1}\left(\frac{C_E}{0.05}\right)^{-1}.
\end{equation}
This means that for massive GC $\beta_{plasma}$ is a constant, 
which depends neither on the specifics of the ICM conditions
including turbulence, nor on the GC mass or age.
It is instead simply determined by two 
fundamental parameters, $C_E$ and $\eta_{turb}$, which describe the efficiency of turbulent 
dynamo and of turbulent heating in structure formation, 
respectively. This is shown in Figure~\ref{ICM_turb_evol:fig}a where $\beta_{plasma}$
is plotted as a function of cosmic time and exhibits 25\% rms fluctuations, which should 
also characterise massive cluster-to-cluster variations.
We can also compute the Alfv\'en scale. Since the turbulence is non-stationary and the magnetic 
energy retains memory over more than one eddy turnover time, we average the dissipation rate 
$\epsilon_{turb}$ over 2~Gyr when calculating $L_A$. Expressing $L_A$ in units of the turbulence outer
scale we write
\begin{equation}
\frac{L_A}{L}\equiv \frac{v_A^3}{C^\frac{3}{2} \langle\epsilon_{turb}\rangle}= 
\frac{3}{2}\left(\frac{2}{\gamma\beta_{plasma}}\right)^{\frac{3}{2}}\frac{c_s^3}{\langle(\delta u_L)^2\rangle^\frac{3}{2}}=
\frac{1}{100}\left(\frac{\beta_{plasma}}
{40}\right)^{-\frac{3}{2}}\left(\frac{{\cal M}_{turb}}{1}\right)^{-3}.
\end{equation}
We have already shown that both $\beta_{plasma}$ and the turbulence 
Mach number, ${\cal M}_{turb}=\langle(\delta u_L)^2\rangle^\frac{1}{2}/c_s$, remain constant during 
the evolution of the GC. Therefore, the Alfv\'en scale too remains a constant fraction of the turbulence 
driving scale, independent of time, GC mass and ICM conditions. In addition, given the large value 
of $\beta_{plasma}$, and that ${\cal M}_{turb}\approx 1$, $L_A$ is small compared to $L$.
The time evolution of $L_A/L$ is shown in Figure~\ref{ICM_turb_evol:fig}b (see Fig.~\ref{ICM_turb_spectrum:fig} in Methods for a typical ICM spectrum of hydromagnetic turbulence).
Finally, Figure~\ref{ICM_turb_evol:fig}c shows that the evolution of the turbulent injection scale, $L$, 
closely follows that of $L_A$ while tracking the growing characteristic scale of the 
GC ($R_{500}=0.5R_{vir}$) also plotted in the same panel.
Note that the modulation of $L_A$ reflects the changing turbulent conditions in the ICM and,
in particular, is anti correlated with $\epsilon_{turb}$, as generally expected.

The large value of $\beta_{plasma}$  indicates that magnetic energy, like turbulent energy, is small 
compared to thermal energy ($E_{th}\gg E_{B}$). Moreover, the small value of $L_A/L$ indicates 
that the dynamo is far from saturation and magnetic energy is also small in comparison to the 
turbulent kinetic energy ($E_{turb}\gg E_B$). 
This energy hierarchy is fundamentally due to the efficiency $\eta_{turb}$ with 
which turbulent energy is generated during gravitational collapse and the fraction $C_E$ thereof
that is converted into magnetic energy, namely $E_{th}:E_{turb}:E_B=1:\eta_{turb}:C_E\eta_{turb}$.
The values of $\beta_{plasma}$ and $L_A$ are in good agreement with recent measurements 
of magnetic field properties in GC~\cite{2001ApJ...547L.111C,2008A&A...483..699G,Bonafede2010,Govoni2010,2011A&A...529A..13K}. 
Here, they emerge from pure numerical modelling of structure formation turbulence and MHD dynamo action, 
in the sense that the are found to derive their values from the parameters $\eta_{turb}$ and $C_E$,
which are determined numerically and not through parametric fits.
Intriguingly, the above energy hierarchy appears to remain unchanged during the GC evolution and the 
turbulent dynamo in the ICM is far from saturation today as it has virtually always been in the past.
Figure~\ref{ICM_turb_evol:fig}c shows that the GC size and, therefore, its mass constantly grow. 
This implies that the gravitational potential energy and therefore the ICM thermal energy and turbulent 
energy also continue to grow. 
Meanwhile dynamo action tries to bring magnetic and turbulent energy into equipartition.
Since all of 
these forms of energy grow simultaneously but with different constant efficiencies, their ratio remains 
unchanged, reflecting the value of those intrinsic efficiencies. In other words, both $\beta_{plasma}$
and $L_A/L$ encode the efficiency of turbulent generation in structure formation and the efficiency of
dynamo action. As such, they allow us to relate magnetic field observations in massive galaxy clusters
to such properties of structure formation. This is in sharp contrast with other astrophysical 
bodies\cite{2008Natur.454..302B,2014Natur.514..597S,2014Natur.510..126Z}, 
e.g. the interstellar medium of galaxies, stars and compact objects, where the turbulence dynamo 
has long saturated and such information is lost forever.

\newpage

\begin{figure}
\includegraphics[width=1.05\columnwidth,angle=0]{fig1.pdf}
\caption{{\bf High resolution simulation of a galaxy cluster in fully cosmological context.}
Baryonic gas in the large scale structure of the universe (panel a;
bright is high, dark is low), and around the GC centre 
where numerical resolution is highest (panel b; colormap inverted). 
Dynamic range of density (in $cm^{-3}$) is $\approx 10^6$.
The black dash-line marks the virial radius, $R_{vir}$, enclosing the volume 
that has nominally reached dynamical equilibrium. 
Panel c: vorticity magnitude on a scale twice the finest mesh size ($\simeq 20$~kpc).
Panel d: temperature map. Complexity is due to shocks and contact discontinuities in the turbulent flow.
Dynamic range of temperature (in K) and vorticity (in $H_0^{-1}$) is about $10^3$.
\label{GC:fig}
}
\end{figure}

\begin{figure}
\includegraphics[width=1\columnwidth]{fig2.pdf}
\caption{{\bf Time evolution of turbulence.}
Panel a: $\eta_{turb}$, the ratio of ICM thermal energy contributed by turbulent dissipation,
$\int_0^t dt\, \rho\epsilon_{turb}$, to the total thermal energy, $E_{th}$. 
Panel b: turbulent Mach number, the ratio of the turbulent rms velocity 
$\langle(\delta u_L)^2\rangle^\frac{1}{2}$ to the sound speed $c_s$.
Panel c: volumetric turbulent dissipation rate, $\epsilon_{turb}$
obtained in\cite{BM15}.
The error bar correspond to the variance of $\epsilon_{turb}$.
All quantities are computed within 1/3 of the virial radius. 
Time in billion-yr is reported on the bottom x-axis and cosmological redshift on the top x-axis.
\label{epsilon}
}
\end{figure}
\begin{figure}
\includegraphics[width=1\columnwidth]{fig3.pdf}
\caption{{\bf Time evolution of magnetic field.}
Panel a: $\beta_{plasma}$, the ratio of ICM thermal to magnetic pressure computed as
$B^2/8\pi=C_E \int^t d\tau \rho\epsilon_{turb}(\tau)$.
Panel b: $L_A/L$, the ratio of Alfv\'en to the turbulent injection scale. 
$L_A(\tau)=v_A^3/[C^\frac{3}{2}\langle\epsilon_{turb}\rangle]$, where 
$v_A=B/\sqrt{4\pi}$ is the Alfv\'en speed
and $\langle\epsilon_{turb}\rangle$ is the turbulent dissipation rate
smoothed over $\tau=2$~Gyr with a Gaussian filter.
Quantities in panels a,b refer to a volume inside $1/3$ the virial radius.
The dash lines show transients to the asymptotic regime for an artificial $t_{start}=4.5$~Gyr.
Panel c: turbulence injection scale $L$ (solid-line) and the characteristic 
cluster size $R_{500}=0.5~R_{vir}$ (dash-line), enclosing a mass over-density of 500. 
Time in billion-yr (bottom x-axis) and cosmological redshift (top x-axis) are reported.
\label{ICM_turb_evol:fig}
}
\end{figure}
\begin{figure}
\includegraphics[width=1\columnwidth]{fig4.pdf}
\caption{{\bf Generation and cascade of ICM hydromagnetic turbulence.}
First the gravitational potential energy is converted into kinetic energy of accretion flows.
These generate shear and shocks which, in addition to heat dissipation,
produce fluid instabilities and baroclynic term, respectively, leading to turbulent flows.
Shocks also accelerate particles through the Fermi I mechanism.
Shocks do not dissipate tangential flows which will either generate turbulence,
shear or shocks or a combination thereof.
The turbulence cascade includes, dissipation of compressible modes at 
weak shocks, conversion of turbulent to magnetic energy via dynamo action,
excitation of plasma waves accelerating relativistic particles through Fermi II mechanism,
and of course viscous dissipation.
\label{ICM_turb_diagram:fig}
}
\end{figure}
\begin{figure}
\includegraphics[width=1\columnwidth]{fig5.pdf}
\caption{{\bf Spectrum of ICM hydormagnetic turbulent cascade.}
Characteristic spectrum of turbulent kinetic energy in the ICM.
Solid and dashed lines correspond to the solenoidal (Kolmogorov-like) 
and the compressional (Burgers-like) velocity field, respectively. 
On the X-axis, from left to right we have marked the virial scale, $R_{vir}$, 
the injection scale, $L$, the Ozmidov's scale, $L_O$, the AlfvŽn scale, $L_A$, 
and Kolmogorov's dissipation scale, $\ell_{diss}$. All quantities are time dependent and
Ozmidov's scale is comparable to the injection scale, so at times 
turbulence in the radial direction could be suppressed by stratification.
\label{ICM_turb_spectrum:fig}
}
\end{figure}


\begin{methods}

\subsection{Numerical model}

The simulation is carried out with {\tt CHARM}, an
Adaptive-Mesh-Refinement cosmological code~\cite{2007JCoPh.227..400M}.
This code uses a directionally un-split variant of the piecewise parabolic method 
for hydrodynamics\cite{1990JCoPh..87..171C},  constrained-transport 
algorithm for solenoidal MHD~\cite{2011ApJS..195....5M},
a time-centred modified symplectic scheme 
for the collision-less dark matter, and solve PoissonÕs equation with a second-order 
accurate discretisation. 
The magnetic field remains negligible throughout, so the calculation
is effectively hydrodynamic. 
For massive galaxy clusters, such as Coma cluster, the ICM cooling time is a few
times the age of the universe\cite{2013A&A...559A..78G}, 
so cooling and baryonic feedback processes are neglected.
Heating of the intergalactic medium through photoionization is also neglected, 
with no consequences whatsoever for the generation of vorticity and turbulence at 
accretion shocks.
We use a concordance $\Lambda$-CDM universe with
normalized (in units of the critical value) total mass density, $\Omega_m=0.2792$,
baryonic mass density, $\Omega_b=0.0462$, vacuum energy density,
$\Omega_\Lambda= 1- \Omega_m= 0.7208$, normalized Hubble constant
$h\equiv H_0/100$ km s$^{-1}$ Mpc$^{-1}$ = 0.701, spectral index of
primordial perturbation, $n_s=0.96$, and rms linear density
fluctuation within a sphere with a comoving radius of 8 $h^{-1}$
Mpc, $\sigma_8=0.817$~\cite{2009ApJS..180..330K}.
The simulated volume has comoving size of $L_{Box}=240\,h^{-1}$ Mpc on a side.
The initial conditions are generated on three refinement levels
with~\texttt{grafic++} (made publicly available by D. Potter). For the coarsest level 
we use 512$^3$ comoving cells, corresponding to a nominal spatial resolution
of 468.75$\,h^{-1}$ comoving kpc and 512$^3$ particles of mass $6.7\times 10^9\,h^{-1}$~$M_\odot$
to represent the collisionless dark matter component.
The additional levels allow for refined initial conditions in the volume where the galaxy cluster forms.
The refinement ratio for both levels is, $n_\mathrm{ref}^\ell\equiv \Delta x_{\ell}/\Delta x_{\ell+1}=2$, $\ell=0,1$.
Each refined level covers 1/8 of the volume of the next coarser level
with a uniform grid of 512$^3$ comoving cells while the dark matter
is represented with 512$^3$ particles.
At the finest level the spatial resolution is $\Delta x=$ 117.2$\,h^{-1}$ comoving kpc
and the particle mass is $10^8\,h^{-1}$ M$_\odot$.
As the Lagrangian volume of the galaxy cluster shrinks under self-gravity, three additional 
uniform grids covering 1/8 of the volume of the next coarser level
are employed with 512$^3$, 1,024$^3$ and 1,024$^3$ comoving cells, respectively, 
and $n_\mathrm{ref}^\ell=2,4,2$, for $\ell=2,3,4$, respectively.
All of them are in place by redshift 1.4, providing a spatial resolution 
of 7.3 h$^{-1}$ comoving kpc in a region of 7.5 h$^{-1}$ Mpc, accommodating the whole 
virial volume of the GC. The ensuing dynamic range of resolved spatial scales is sufficiently large for the 
emergence of  turbulence.

\subsection{Galaxy cluster characteristic quantities}
The galaxy cluster and its formation history 
are reconstructed using our implementation of a HOP halo finder\cite{1998ApJ...498..137E}
and merger history code.
The virial radius is defined as the region enclosing a mass over-density $\Delta_c =
178\Omega_m^{0.45}$ with respect to the critical density\cite{2001ApJ...554..114E}. 
At redshift $z=0$, the viral radius is $R_{vir} = 1.95 h^{-1}$ Mpc, 
and the corresponding enclosed mass, $M_{vir} = 1.27\times 10^{15} M_\odot$. 
Also a $z=0$, using $\Delta_c=500$ we find the characteristic radius $R_{500}\simeq 1 h^{-1}$ Mpc.

\subsection{Turbulence characterization}
The characteristic quantities describing the turbulence are inferred from the analysis of the
structure functions. This analysis is described in detail in\cite{2014ApJ...782...21M,2015ApJ...800...60M}.
Basically, we decompose the velocity into a solenoidal and a compressional component using
a Hodge-Helmoltz decomposition, i.e.
\begin{equation} \label{hh0:eq}
\mathbf{v}=\mathbf{v}_s+\mathbf{v}_c, \\
\mathbf{v}_c=-\nabla\phi, \quad \mathbf{v}_s=\nabla\times \mathbf{A}, \\
\phi =\frac{1}{4\pi}\int \frac{\nabla\cdot\mathbf{v}}{r} d\mathbf{x}, \quad
\mathbf{A} =\frac{1}{4\pi}\int \frac{\nabla\times\mathbf{v}}{r} d\mathbf{x}. \label{hh2:eq},
\end{equation}
and then we compute the second and third order structure functions
of velocity increments of the solenoidal component,
$\delta{v}_i\equiv [\mathbf{v}_s(\mathbf{x}+\mathbf{l})-\mathbf{v}_s(\mathbf{x})]_i$
\cite{Landau:111625},
\begin{equation}\label{gcf:eq}
S_{i}({\bf l}) \equiv \left\langle (\delta{v}_i)^p \right\rangle,
\end{equation}
where $p=2,3$ indicates the structure function order, and
$i$ indicates the projection along or perpendicular to ${\bf l}$ for the longitudinal 
and transverse structure functions respectively.
To compute the structure functions
we define sampling points randomly distributed inside the volume of interest (within
(1/3) of the viral radius), and 
compute the velocity difference with respect to other randomly selected field points at a
maximum distance of two virial radii.
Once we the velocity structure functions are computed, we define the 
velocity dispersion as the asymptotic values of the second order structure function,
and the outer scale as the separation at which that asymptotic value is reached. 
To compute the Mach number we divide the turbulent velocity dispersions 
by the sound speed, $c_s=\sqrt{\gamma P/\rho}$, computed by evaluating the mean value of
each thermodynamic quantity within the same volume in which the sampling points are collected.
Finally, the turbulent dissipation rate is computed by identifying the inertial range of the second 
and third order structure functions of the solenoidal velocity increments (for details see ref.~\cite{BM15}).

\subsection{Code Availability}
We have opted not to make the code available for practical reasons.
However, the methods we adopt are published in the literature and are commonly used in the community. 
Amongst others, the publications mentioned in the above Methods section
contain tests of our code against problems with known 
solutions and also with respect to solutions obtained with similar codes from independent authors.

\end{methods}

\bibliography{miniati,beresnyak}

\begin{addendum}
 \item This work was supported by a grant from the Swiss National Supercomputing Center (CSCS) under project ID S419 and S506.
 \item[Author Contributions] F.M. carried out the cosmological simulations, computed the turbulence structure functions, derived Eq. 1, 2 and 3 and wrote most of the text. A.B. analysed the structure functions, testing the
 self-similar nature of 2nd and 3rd order structure function within the inertial range and computing the dissipation rate. A.B. and F.M. computed the evolution of $E_B$ and $L_A$.
 \item[Competing Interests] The authors declare that they have no competing financial interests.
 \item[Correspondence] Correspondence and requests for materials
should be addressed to F.M.~(email: fm@phys.ethz.ch).
\end{addendum}


\end{document}